# 2

# Locality and Utilization in Placement Suboptimality

**Jason Cong,**[1] **Michalis Romesis**[2]**, Joseph R. Shinnerl**[3]**, Kenton Sze**[4]**, and Min Xie**[1]

[1]UCLA Computer Science
[2]Magma Design Automation, Inc.
[3]Tabula, Inc.
[4]UCLA Mathematics
`{cong,xie}@cs.ucla.edu`
`michalis@magma-da.com`
`jshinnerl@tabula.com`
`nksze@math.ucla.edu`

## 2.1 Introduction

Placement is a critical step in VLSI design. Interconnect delay dominates system performance, and placement determines the interconnect more than any other step in physical design. The complexity of modern designs, however, makes estimation of suboptimality difficult [14, 16, 28]. Studies on simplified, synthetic benchmarks with known optimal-wire length placements (PEKO [7]) initially suggested that many leading tools may produce solutions with excess wire length from 60% up to 150% or more. These results have generated wide interest in both industry [13] and academia [19, 22, 28]. Recent progress in placement [1, 5, 6, 17] has reduced the wire length gap on PEKO to about 12–40%.

The PEKO benchmarks, however, have well-known limitations. Although their cell counts, net counts, and net-degree statistics match corresponding quantities in standard industrial benchmarks [2], the PEKO circuits are simplified in three key ways, in order to guarantee known optimal solutions. First, all cells are squares of the same size. Second, the known optimal placements for the PEKO circuits are packed layouts with zero white space. Third, all nets in an optimal PEKO placement are

[1]Version of Record: https://link.springer.com/book/10.1007/978-0-387-68739-1
[2]Acknowledgment: This is a pre-print of Chapter 2 from the following work: Gi-Joon Nam and Jason Cong, Modern Circuit Placement Best Practices and Results, 2007, Springer, reproduced with permission of Springer Science+ Business Media, LLC. The final authenticated version is available online at: http://dx.doi.org/10.1007/978-0-387-68739-1

local – the netlist of a PEKO circuit is defined over cells arranged in a regular array, with adjacent cells grouped into local nets of minimum HPWL.

Subsequent studies [9, 16] derived useful lower bounds on the HPWL suboptimality of placements of circuits with more realistic netlists. The PEKU circuits [9] add nonlocal nets to packed, uniform-grid PEKO layouts but sacrifice any assurance of optimality. Zero-change netlist transformations [16] preserve both module shapes and core utilization, but they quantify the sensitivity of a placement tool to netlist changes, not the suboptimality of a given placement on a given netlist. It is not known how close the lower bounds on suboptimality are to the true suboptimality gaps for either the PEKU circuits or the zero-change netlists.

The benchmarks described in this chapter directly address several of the shortcomings in existing suboptimality benchmarks. Two new sets of placement examples are constructed, one targeting the role of nonlocal nets in suboptimality, and another targeting the role of white space and large variations in module sizes. The first set, PEKO-MC, is a set of standard-cell circuits with nonlocal nets in known optimal placements. A given netlist is modified so as to render a given placement for it optimal for the new netlist. Cell dimensions and locations are not changed, net-degree statistics are matched exactly, and over 60% of the original netlist is left unchanged. The second set, PEKO-MS, incorporates a parametrized percentage of white space into a mixed-size placement which precisely matches given macro dimensions and locations as well as the net-degree distributions of the ISPD 2005 benchmark suite [21]. HPWL for the placements generated for the PEKO-MS circuits are proven to be less than 3% above optimal for most cases and within 8% of optimal on all cases.

The concept of a monotone path for a circuit signal has been used in performance-driven logic synthesis [23, 26], coupled timing-driven placement and logic synthesis [27], performance-driven multilevel partitioning [15], and the analysis of wire length models in timing-driven placement [24]. The concept is also employed in two of the three netlist transformations used by Kahng and Reda [16]. To our knowledge, however, the work described here is the first to employ monotone chains in the construction of netlists with known optimal-wire length placements.

Typically, mixed size placement proceeds in three stages: global placement (GP), legalization, and detailed placement (DP). The goal of *GP* is to position each cell within some relatively small neighborhood of its final position, while eventually obtaining a sufficiently uniform distribution of cell area over the entire chip. Typically, large sets of cells are moved simultaneously under some relaxed or incremental formulation of area density control – scalable algorithms do not strictly enforce pairwise nonoverlap constraints during this stage. The goal of *legalization* is, given a sufficiently good GP $P_g$, determine positions of all cells so that (1) no two cells overlap and (2) a given objective, e.g., approximate total wire length or total displacement from $P_g$, is minimized. During *DP*, all constraints are strictly enforced. Typically, DP proceeds by a sequence of refinements made one at a time on small, contiguous subregions [4, 12] or on individual rows [18].

In practice, GP is terminated when iterations are observed to make little or no reduction in the objective and the module-area distribution is sufficiently uniform.

How much of the optimality gap left by contemporary methods should be attributed to deficiencies in global-placement algorithms, and how much to legalization and DP? On a real circuit, there is no way of knowing how far a cell is from its nearest optimal location, at any stage. On circuits with known optimal or near-optimal placements, however, it is possible to evaluate precisely the quality of any of the three engines in isolation from its counterparts. Thus, the benchmark circuits described here provide a more precise means of quantifying the relative effectiveness of the methods used in the three stages. Results estimating the separate suboptimality contributions of GP and legalization and DP are described in Sect. 2.4.

## 2.2 Peko-MC Benchmark Construction

Each PEKO-MC example has an optimal-wirelength placement in which over 50% of the nets are non-local. Module shapes, core utilization, and net-degree statistics match corresponding quantities in a given benchmark exactly. The PEKO-MC construction is described in this section.

### 2.2.1 Monotone Chains

The definition of a monotone chain in a netlist uses simple ideas from both graphs and hypergraphs. First, consider a path $P$ in a *graph G* whose vertices lie in the plane. Let $P$ consist of $n$ consecutive edges $(e_1, \ldots, e_n)$ connecting $n + 1$ vertices $(v_0, \ldots, v_n)$, vertex $v_i$ with coordinates $(x_i, y_i)$ and edge $e_i$ connecting vertices $v_{i-1}$ and $v_i$. Then $P$ is *monotone* if and only if, for every $i \in \{1, \ldots, n\}$, $|x_n - x_i| \leq |x_n - x_{i-1}|$ and $|y_n - y_i| \leq |y_n - y_{i-1}|$. Hence, a path in a graph embedded in the plane is monotone if and only if the Manhattan distance between its two terminal vertices equals the sum of the Manhattan lengths of its edges.

In a hypergraph, a *hyperpath* is a finite sequence of hyperedges in which each hyperedge intersects with its predecessor and successor. We say that a hypergraph lies in the plane or, equivalently, is placed in the plane, if the nodes (modules) of the hypergraph (netlist) have been assigned specific locations in the plane. In this case, the total length of a hyperpath is the sum of the HPWLs of its hyperedges (HPWL denotes minimum bounding-box half-perimeter). An edge $e = (v, w)$ is called the *equivalent edge* of a hyperedge $h$ of a hypergraph in the plane, if (1) its vertices $v$ and $w$ are in $h$ and (2) $e$'s minimum-HPWL bounding box is the same as $h$'s minimum-HPWL bounding box. A hyperedge in the plane may have zero, one, or two equivalent edge(s).

A path $P$ is called the *equivalent path* of a hyperpath $H$ in a hypergraph in the plane, if there is a one-to-one correspondence between the edges of $P$ and the hyperedges of $H$, such that every edge of $P$ is the equivalent edge of its corresponding hyperedge in $H$. A *monotone chain* is a hyperpath which has an equivalent path that is monotone.

Assuming that no two vertices can occupy the same location, neighboring hyperedges in a monotone chain have exactly one vertex in common. These common vertices form the equivalent monotone path. The two terminal vertices of a monotone chain are the terminal vertices of its equivalent path. Hence, the length of a monotone chain equals the HPWL of the edge defined by the chain's two terminals.

**Observation 2.2.1** *If the terminal vertices of a monotone chain* $P = (h_1, h_2, \ldots, h_n)$ *of a hypergraph G are fixed in the plane, there is no other planar embedding of hypergraph G which reduces the length of P.*

Given a placement of the hypergraph in the plane, a *local net* is a hyperedge the HPWL of which is the minimum possible, subject to some spacing constraints between vertices. From Observation 2.2.1, it is evident that a placement has optimal HPWL if all its nonlocal nets can be partitioned into netwise-disjoint monotone chains with fixed endpoints.

### 2.2.2  The Peko-MC Algorithm

Starting from the placement of the real benchmark, sets of nets are identified that can be grouped together into netwise-disjoint monotone chains between well-separated fixed terminals. Initially, these chains are not complete and have gaps called intervening regions. These are later filled by other nets that are modified from the original netlist. Local nets in the given placement are not modified. The main steps of the PEKO-MC algorithm are sketched below.

*Placement generation*. The PEKO-MC generator requires a placement of the original netlist. This placement is held fixed, while the netlist is changed so that the given placement attains the optimal HPWL for the modified netlist. Starting from a random placement is possible, but experiments show that starting from a placement computed by a tool increases the final similarity of the original and the derived circuits, because a real placement has many more locally optimal nets than a random placement.

*Net categorization*. The nets of the original hypergraph are divided into three different categories depending on the placement of their pins:

(1)  Locally optimal-HPWL nets
(2)  Nets that do not have equivalent edges
(3)  Nets with equivalent edges

Nets of Type (2) cannot be members of monotone chains and are therefore modified. Nets of Type (3) are labeled according to the directions of the monotone chains of which they can be members: from lower left toward upper right, or from lower right to upper left. Some of these nets can be members of chains in either direction.

*Chain generation*. As illustrated in Figure 2.1, sets of nets that can be members of the same chain are identified along with sets of intervening regions that must later be filled by nets in order to complete the monotone chains. All nets of Type (3) are assigned to chains during this step.

*Chain removal*. In our experiments, the number of intervening regions between pairs of nets created during chain generation is higher than the number of nets of Type (2). Hence, to preserve netlist statistics, some of the chains generated are removed in order to reduce the number of such intervening regions and increase the number of available nets. Chains with the highest ratios of intervening regions to contained nets are removed until the number of available nets equals or exceeds the number of gaps between terminals in chains.

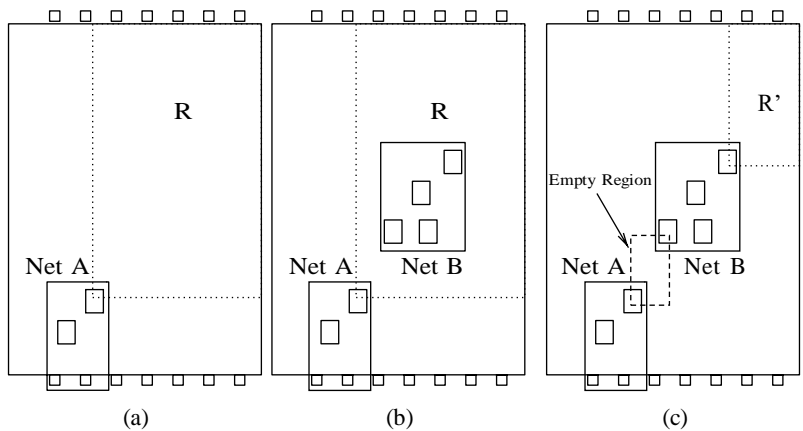

**Fig. 2.1.** Example of chain generation. (**a**) Net A, containing two cells, has already been added to the chain. A search for a new net takes place in region R. (**b**) Net B is selected to be added to the chain. (**c**) An intervening region is inserted between nets A and B that will be covered later by a new net. A new search is initiated for nets in region R'.

*Gap covering.* In the final step, empty regions between nets in chains are filled by new nets. Each new net replaces some available net in the original netlist. The new net includes the two pins defining the equivalent edge of the bounding box of its intervening region *R* as well as additional pins selected from within *R* in order to match the degree of the replaced net. The cells whose degree in the current netlist are smallest compared to their original degree are given priority. In this way, the cell-degree distribution of the new netlist closely follows the corresponding distribution of the original circuit. Most intervening regions are covered by one net, but a few are covered by two nets, when the number of available nets exceeds the number of empty regions.

Experiments reported in Sect. 2.4 suggest that, on the 2004 FastPlace-IBM standard-cell circuits with 20% white space, nonlocal nets probably do not represent a significant source of suboptimality for these tools. In order to amplify the suboptimality observed on mixed-size cases as much as possible, the PEKO-MS benchmarks described next include only local nets by default.

## 2.3 Peko-MS Benchmark Construction

We refer to our placement suboptimality benchmarks with parametrized white space as PEKO-MS. As shown in Figure 2.2, the PEKO-MS generator produces a benchmark closely approximating the following four targets (1) net-degree histogram $N_\#$, (2) given placement $P_{mac}$ of all macros, (3) number of standard cells $N_{sc}$, and (4) white

Set grid-resolution limit $\overline{N_G}$.

**input**

| | |
|---|---|
| $N_\#$ | target net-degree histogram |
| $P_{\text{mac}}$ | macro placement in core region $R$ |
| $N_{\text{sc}}$ | target number of standard cells |
| $\varphi_{\text{ws}}$ | target white-space fraction |

$\varphi_{\text{mac}} := \left( \sum_{v_i \in P_{\text{mac}}} a(v_i) \right) / a(R)$.

$\varphi_{\text{ws}} := \min\{\varphi_{\text{ws}}, 1 - \varphi_{\text{mac}} - N_{\text{sc}}/N_G\}$.

$\varphi_{\text{sc}} := 1 - \varphi_{\text{mac}} - \varphi_{\text{ws}}$.

$N_G := N_{\text{sc}}/\varphi_{\text{sc}}$.

**if** $(N_G > \overline{N_G})$ **then**

   $N_G := \overline{N_G}; \varphi_{\text{sc}} := N_{\text{sc}}/N_G; \varphi_{\text{ws}} := 1 - \varphi_{\text{mac}} - \varphi_{\text{sc}}$.

**end if**

Snap $P_{\text{mac}}$ into $G$, truncating macros as necessary;

   mark grid cells assigned to macros.

$N_{\text{ws}} := \varphi_{\text{ws}} \cdot N_G$.

**repeat**

   Randomly select unvisited non-macro grid cell $c$.

   **if** (the spatial neighbors of $c$ remain spatially

      connected in $G$ when $c$ is removed) **then**

      Mark $c$ as white space and decrement $N_{\text{ws}}$.

   **end if**

**until** ($N_{\text{ws}} == 0$ **or**

         every non-macro grid cell has been examined)

**if** ($N_{\text{ws}} > 0$)   **report failure and exit**   **end if**

Mark all unmarked grid cells as standard cells;

   $V := \{\text{macros}\} \cup \{\text{standard cells}\}$.

Following Figure 2.3, generate a minimal netlist

   "backbone" $E_B$, a connected set of local nets

   consistent with $N_\#$ which covers $V$.

**while** ($N_\#$ still has nonzero entries **and**

         available locations for local nets still exist)

   Randomly select an available location $p$ for a local net

   **if** (no new local net can be generated at $p$) **then**

      **remove** $p$ from list of available local-net locations.

   **else**

      generate a local net of maximum possible degree $k$

      still represented in $N_\#$. Decrement $N_\#[k]$.

   **end if**

**end while**

**output** the placement suboptimality benchmark netlist space fraction $\varphi_{ws}$. The $i$ th component of vector $N_\#$ is the target number of nets of cardinality $i$. A macro is any module, fixed or movable, with height greater than the standard-cell row height. The generator places $N_{sc}$ standard cells between macros and defines nets locally such that the total HPWL of the given placement is no more than a small, explicitly computed factor (1.00–1.08) above optimal for the final benchmark. Connectivity of the constructed netlist is ensured by inserting white space in such a way that all remaining cells and macros form a spatially connected set in the placement region.

**Fig. 2.2.** The Peko-MS benchmark generator.

As described in Figure 2.2 and later, the PEKO-MS generator proceeds in four stages:

1. Input target statistics; definition of uniform grid $G$; definition of mapping $f_G$ which snaps a given macro placement $P_{mac}$ into $G$.
2. Designation of white-space grid-cells, leaving cells, and macros spatially connected.
3. Construction of the netlist backbone (Figure 2.3), a minimal connected set of local, near-optimal-HPWL nets connecting all cells and macros.
4. Construction of additional, optimal-HPWL local nets to match target netlist statistics as closely as possible.

An optional additional stage for the addition of optimal-HPWL *nonlocal* nets is described in Sect. 2.4.

Every legal mixed-size placement induces a complicated partition $R = R_{mac} \cup R_{sc} \cup R_{ws}$ of its placement region $R$ into three disconnected subregions: $R_{mac}$ occupied by macros, $R_{sc}$ by standard cells, and $R_{ws}$ left as white space. The PEKO-MS generator preserves a given macro placement $P_{mac}$ precisely with respect to a fixed core region $R$. Let $a(S)$ denote the area of subregion $S$. Region $R$ is neither shrunk nor expanded relative to the macros – both $a(R)$ and $a(R_{mac})$ are held fixed. Instead, standard cells are uniformly shrunk or inflated to attain a higher or lower white-space targets, respectively. With this fixed-outline and fixed-macro-layout strategy, $\varphi_{mac} \equiv a(R_{mac})/a(R)$ is fixed, and it is evident that white space cannot be increased beyond the space left to it by the macros and standard cells:

$$\varphi_{ws} \leq 1 - \varphi_{mac} - \varphi_{sc}^{min}$$

where $\varphi_{sc}^{min}$ denotes the minimum fraction of $R$ which can be left for standard cells. The exact value of $\varphi_{sc}^{min}$ is determined by storage and run-time considerations, as described next.

A tight lower bound on the optimal HPWL of each PEKO-MS benchmark is obtained by mapping the given macro layout $P_{mac}$ into a uniform rectangular integer grid $G$ of square cells over which all nets are defined. The mapping is denoted by $f_G P_{mac} \rightarrow \text{Rect}(2^G)$, where $\text{Rect}(2^G)$ denotes the set of all contiguous rectangular subsets of grid cells in $G$. Each macro is identified by the mapping $f_G$ with a distinct rectangular subset of grid cells in $G$. A nonoverlapping macro placement ensures that the grid-cell subsets associated with distinct macros are disjoint. Each center of each grid cell represents a candidate pin location. Pin locations on macros

---

**input**    $C := \varnothing$ = set of vertices contained in nets
$B := \varnothing$ = set of vertices not yet in $C$ but spatially
        adjacent (in $G$) to at least one vertex in $C$.
Create a local net $e$ at a random location.
Insert all $v \in e$ into $C$ and all $G$-neighbors of $e$ into $B$.
**while** ($B$ is not empty)
      Select an as yet unconnected grid cell $b \in B$ and a
          connected grid cell $c \in C$ such that $b$ and $c$ are
          adjacent in $G$. Cell $b$ may be either a standard cell
          or a grid cell assigned to the boundary of an as yet
          unconnected macro. Cell $c$ may be either a
          standard cell or an as yet unconnected grid cell
          assigned to the boundary of a connected macro.
      Create a net $e$ containing $b$ and $c$ and containing as
          many other standard cells as possible, up to the
          maximum target net degree remaining in $N_\#$.
      **if** ($N_\#[|e|] > 0$) **then** decrement $N_\#[|e|]$
      **else**
          $k := \min\{j \mid j > |e| \text{ and } N_\#[j] > 0\}$.
          decrement $N_\#[k]$ and increment $N_\#[k - |e|]$,
      **end if**
      **for** ( all $v \in e$ )
          **remove** $v$ from $B$ and insert it into $C$.
          **for** ( each grid neighbor $w$ of $v$ )
              **if** ($w \not\in e$ and $w \not\in C$) **insert** $w$ into $B$ **end if**
          **end for**
      **end for**
**end while**
**output** minimal connected netlist $E_B$ covering all $v \in V$

---

**Fig. 2.3.** Peko-MS Netlist backbone generator.

are restricted to grid-cells on macro boundaries and kept distinct. I.e., the center of each grid-cell along any macro's boundary can serve as a pin for at most one net. For simplicity, however, all pins on each standard cell are located at the same point at the center of that cell; i.e., the center of each standard cell may represent several pins for several different nets.

With all $t$ pins of a given net placed at distinct grid-cell centers, the minimum HPWL of a $t$-pin net in such a grid is $r+s-2$, where $r = \lceil \sqrt{t} \rceil$ and $s = \lceil t/r \rceil$. This result is easily derived by packing the $t$ square grid cells of the net into a rectangle of least possible perimeter. However, as shown in Figure 2.4, the optimal HPWL for a $t$-pin net may be attained by pin configurations with bounding boxes of different shapes.

In order to construct a local net of optimal or near optimal HPWL containing a small subset of rectilinearly connected seed pin locations, rectangles of gradually increasing sizes containing the seeds are recursively examined. Each such rectangle is a rectangular subset of grid-cells containing the seed locations and representing

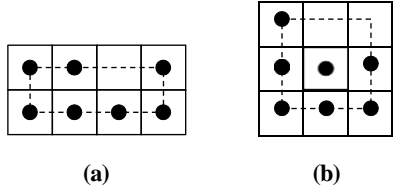

**(a)**                    **(b)**

**Fig. 2.4.** On a uniform square grid, the optimal HPWL of a 7-pin net (4 grid units) can be attained by pin configurations with either of the two bounding boxes shown as *dashed line* segments.

a bounding box for a candidate net. In addition to the seeds, it may contain white space, standard cells, or grid-cells on the boundaries or interiors of macros. Of these, the available pin locations are the centers of the standard cells and the centers of the grid-cells on macro boundaries which have not yet been used as pins in other nets. As long as the number of available, rectilinearly connected pin locations in each such rectangle $R$ is high enough to ensure optimal HPWL of the corresponding net, four larger rectangles containing $R$ may also be considered. As shown in Figure 2.5, a rectangle is enlarged by adding to it a row or column of grid-cells along one of its four edges. Hence, the candidate rectangles for a given set of seeds form a quad-tree, the rectangles increasing in size along any path from root to leaf. Rectangles are enlarged until either optimal-HPWL cannot be obtained or the maximum-degree net remaining in $N_\#$ can be formed.[1]

At each seed location, the highest-degree optimal-HPWL net possible is formed, subject to the constraint that the number of nets of that degree in $N_\#$ has not yet been attained in the benchmark. The reason to form high-degree nets first is simply that they are the most difficult to construct. As pin locations along macros are gradually taken, high-degree nets become ever harder to construct. As not all high-degree targets in $N_\#$ may be attained during the construction, a compromise is made in the backbone-construction phase. When the degree $d_b$ of a large backbone net $b$ is no longer available in $N_\#$ but a larger target degree $d_t > d_b$ in $N_\#$ exists (i.e., $N_\#[d_t] > 0$ for some (net-degree) index $d_t > d_b$), then (1) net $b$ is retained in the constructed netlist, (2) the maximum net-degree target remaining in $N_\#$ is decremented, and (3) the difference degree target entry $N_\#[d_t - d_b]$ is incremented. In this way, the total pin count of the constructed netlist is typically assured of matching the total pin count in the original benchmark.

As is suggested by the labeling in Figure 2.5, incremental enumeration of distinct candidate optimal-HPWL bounding boxes amounts to the enumeration of distinct finite sequences $\{d_i\}_1^N$, where each $d_i \in \{n, s, e, w\}$ represents the direction of enlargement at the $i$th step, and $N = 1, 2, \ldots$ is the total number of enlargements for a given box. Two sequences of the same length $N$ are distinct if and only if the numbers of occurrences of all the symbols $\{n, s, e, w\}$ are not the same for both.

---

[1] To reduce search time, rectangles after a certain level in the quad-tree are enlarged in only one of the most promising directions, i.e., a direction containing the most available pin locations.

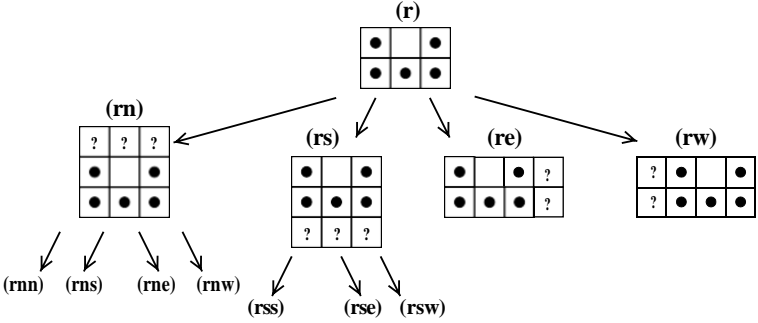

**Fig. 2.5.** The first level of local search for the largest optimal-HPWL net containing a given 5-pin seed. After the first level, many duplicate (e.g., *rsn*) and suboptimal (e.g., *ree*) cases at the subsequent levels can be pruned.

E.g., *ns* and *sn* are equivalent and lead to the same bounding box containing the initial seed box, but *nse* and *nsn* are distinct. The number of distinct sequences of length $N$ is the number of ways $p_4(N)$ that the integer $N$ can be expressed as the sum of four non-negative integers; asymptotically, $p_4(N)$ grows with order $N^3/6$.[2] However, sequences for suboptimal bounding boxes (such as *ree* and *rew* in Figure 2.5) and their descendants can be easily avoided.

Ideally, the resolution of grid $G$ should be high enough to capture all macro and cell dimensions exactly. Our implementation simplifies the definition of $f_G$ in two ways. First, $P_{mac}$ is represented in floating point; macro positions and dimensions are expressed as fractions of chip dimensions prior to their conversion to integer grid units. Macro dimensions are truncated in $G$ as needed to snap macros into the grid.[3] Second, each standard cell is represented by just one of the square grid cells of $G$ – variations in standard-cell width are ignored. These two assumptions significantly reduce the size of $G$ necessary to accurately represent $P_{mac}$. However, the resolution of $G$ must still be large enough that:

(1) Each macro has nonzero height and width.

(2) The number of grid cells not used for macros is large enough to form both the requested number of standard cells $N_{sc}$ and the requested fraction of white space $\varphi_{ws}$.

## 2.4 Experiments

Four sets of experiments with leading academic placement tools are reported. The first is on standard-cell PEKO-MC circuits generated from the 2004 FastPlace-IBM benchmarks. The second is on mixed-size PEKO-MS circuits derived from the ISPD 2005 suite. The third considers the impact of introducing chains of optimal-HPWL nets into a PEKO-MS benchmark. The fourth examines the suboptimality of legalization and detailed-placement engines in isolation from their global-placement counterparts on a parametrized adaptation of the PEKO-MS circuits.

---

[2] The precise expression is $p_4(N) = (N^3 + 6N^2 + 11N + 6)/6$, which is the coefficient of $x^N$ in the Taylor series for $(1 - x)^{-4} = (1 + x + x^2 + x^3 + \cdots)^4$, assuming $|x| < 1$ [3].

[3] A small fraction of the transformed macros in $G$ may be discarded due to error incurred in the truncation, e.g., macros mapped to zero-width rectangles in $G$, or one of a pair abutting macros in $P_{mac}$ which overlap in $G$.

### 2.4.1 Nonlocal Nets (Peko-MC)

All PEKO-MC benchmarks used in our experiments are generated from the Fast-Place [8] versions of the 2002 IBM/ISPD benchmarks [2]. The white space in these test cases is approximately 20%. The FastPlace-IBM benchmarks modify the original IBM benchmarks by replacing macros with standard cells. However, the PEKO-MC algorithm can also be applied to examples with macros for the generation of mixed-size circuits with known optimal placements. Although no new pads are explicitly inserted, most existing pads are connected to several nets each to allow for more chains.

The PEKO-MC benchmark generator described in Sect. 2.2 requires as input both a netlist and an initial "seed" placement of that netlist. Dragon 3.01 [25] and mPL4 [10] were used to seed separate suites of PEKO-MC benchmarks. Using other placers as seeds was observed to have negligible impact on final results, even when the placer used to create the seed placements was run on the resulting PEKO-MC netlists.

The PEKO-MC suite matches the FastPlace-IBM benchmarks exactly in number of cells, cell areas, number of nets, and net-degree distribution. Roughly 60–70% of the nets in the original and synthetic benchmarks are identical, and the distributions of net lengths in the optimal placement of the synthetic benchmarks are nearly identical to those of their seed placements on the original netlists. Moreover, the cell-degree distributions of the original and synthetic benchmarks are very similar (the *degree* of a cell is the number of nets containing the cell). Almost 80% of the cells in an PEKO-MC netlist have a cell-degree difference at most 1 from their corresponding cells in the original netlist. Detailed statistics are shown in Figures 2.6 and 2.7.

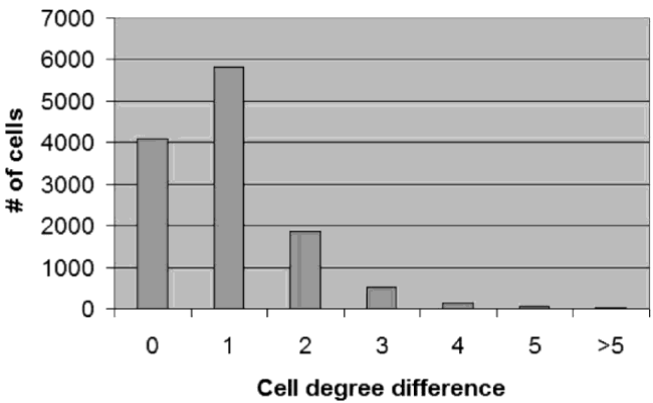

**Fig. 2.6.** The cell-degree difference (in absolute values) distribution between the cells of mPL-MC01 and their corresponding cells in FastPlace-ibm01.

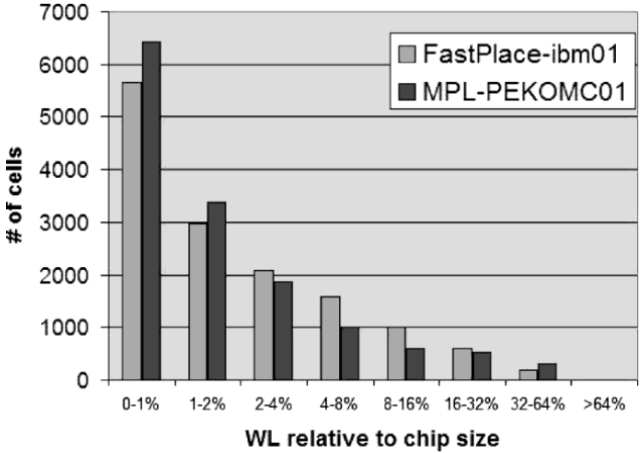

**Fig. 2.7.** The wire length distribution (relative to the chip half-perimeter) of the nets in FastPlace-ibm01 (as placed by mPL4) and the nets in mPL-MC01 (in their optimal placements).

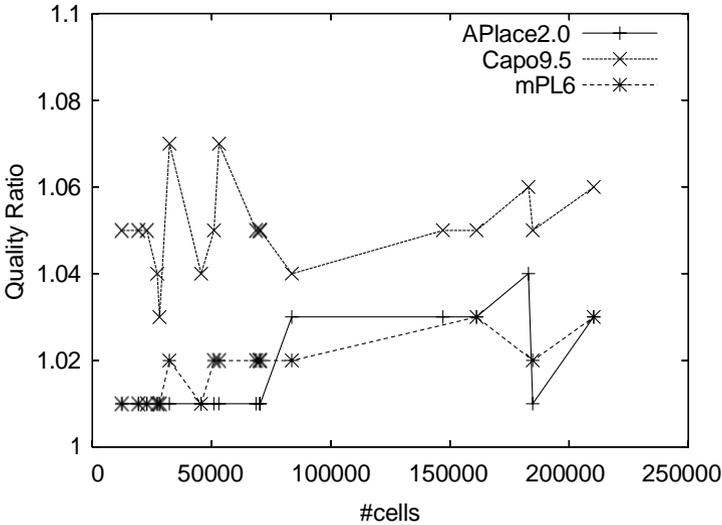

**Fig. 2.8.** Results of some leading academic tools on MC Circuits seeded by Dragon 3.01 placements of FastPlace-IBM Benchmarks.

Results for programs APlace 2.0 [17], mPL6 [6, 11], and Capo 9.5 [1] on the Dragon-MC suite are shown in Figure 2.8. Very similar results (not shown) were obtained for all the tools on the mPL4-MC suite. The overall results show very good performance by all tools on all the benchmarks, regardless of which tool generates the initial placement used to seed the benchmark construction. The worst reported quality ratio by any of the placers on any benchmark is 1.07. We attribute this result

**Table 2.1.** Peko-MS benchmark circuit statistics, with notation.

| circuit | $\overline{N_{sc}}$ | $N_{mac}$ | $\overline{N_{nets}}$ | $\varphi_{mac}$ | $\varphi_{ws}^0$ | $\varphi_{ws}^{max}$ | $\rho_{max}$ |
|---|---|---|---|---|---|---|---|
| PWS-A1 | 216180 | 63 | 233982 | 0.43 | 0.24 | 0.30 | 1.01 |
| PWS-A2 | 264793 | 159 | 299358 | 0.62 | 0.21 | 0.29 | 1.03 |
| PWS-A3 | 474287 | 723 | 531843 | 0.62 | 0.25 | 0.28 | 1.03 |
| PWS-A4 | 531245 | 1329 | 563521 | 0.49 | 0.37 | 0.38 | 1.03 |
| PWS-B1 | 280141 | 32 | 301577 | 0.17 | 0.46 | 0.46 | 1.01 |
| PWS-B2 | 583514 | 23084 | 624625 | 0.38 | 0.38 | 0.40 | 1.03 |
| PWS-B3 | 1137839 | 3778 | 1265913 | 0.67 | 0.14 | 0.24 | 1.08 |
| PWS-B4 | 2237605 | 8170 | 2469988 | 0.38 | 0.35 | 0.40 | 1.03 |

| | |
|---|---|
| $\overline{N_{sc}}$ | average number of standard cells |
| $N_{mac}$ | number of macros |
| $\overline{N_{nets}}$ | average number of nets |
| $\varphi_{mac}$ | macro-area utilization |
| $\varphi_{ws}^0$ | original benchmark's white-space fraction |
| $\varphi_{ws}^{max}$ | maximum white-space fraction attained by the generator |
| $\rho_{max}$ | maximum ratio of generated HPWL to its lower bound |

Averages and maxima are taken over the 4 different white-space values by which each circuit is parametrized. Standard deviations of $N_{sc}$ range from 0.5% to 4.2% of $N_{sc}$; standard deviations of $\overline{N_{nets}}$ range from 10% to 16% of $\overline{N_{nets}}$.

to the increased range of optimal locations available to modules in multipin nets of monotone chains.

### 2.4.2 Parametrized White Space (Peko-MS)

The PEKO-MS approach gives the user control over the layout of the macros. In the ISPD 2005 benchmarks, all macro locations are prespecified for all circuits anyway, except `bigblue3`. For our construction based on `bigblue3`, we extracted movable macro locations from the placement generated for it by APlace [17] for the ISPD 2005 placement contest [29].

Each PEKO-MS local net's construction proceeds by depth-limited local search from a given subset of adjacent grid cells. A small amount of HPWL suboptimality is tolerated in some nets to simplify the implementation.[4] The optimal and attained HPWLs of the individual nets are simply added up to determine the limit on the total HPWL suboptimality in the final benchmark. These limits are shown in Table 2.1. On some circuits, nets $e$ in the source netlist with more than a few hundred pins are represented by small subsets of high-degree nets whose pin counts sum to $|e|$.

Quality ratios of mPL6, APlace 2.0, and Capo 9.5 are listed in Table 2.2. The results show substantial variation both between tools and across different white-space values.

### 2.4.3 Suboptimality Under Both Parametrized White Space and Nonlocal Nets

The preceding results separate the impact of white space and mixed-size modules from that of nonlocal nets. However, the PEKO-MC and PEKO-MS techniques can be combined into a single set of suboptimality benchmarks supporting parametrized percentages of both nonlocal nets and white space. A combination derived from the Peko-MS construction (Figure 2.2) was tested on the mixed-size IBM01 benchmark from the ICCAD2004 test suite [1], as follows. Following the construction of the Peko-MS netlist backbone (Figure 2.3), monotone chains of nonlocal nets are constructed as follows:

1. The set of all boundary pads and candidate pin locations of fixed macros is partitioned by a simple heuristic into pairs of fixed terminals, such that the terminals in each pair are relatively far apart.

2. For each pair of terminals, designate one terminal in the pair as the start, and another as the end. A chain of nonlocal nets is iteratively constructed for the pair of terminals by the following sequence of steps (compare to Figure 2.1):

   (a) Randomly select an available pin location in the bounding box of the end terminal and the net corner pin most recently added to the chain. The selected location is the next net corner pin.

   (b) Randomly select additional pins in the resulting bounding box of that new net-corner pin location and the preceding net-box corner pin to populate the net.

---

[4] However, we still refer to the placements as optimal, because the set of modules in each net is rectilinearly connected and hence supports an optimal *routed* wire length of the net.

**Table 2.2.** Results for mPL6, APlace 2.0, and Capo 9.5 on Peko-MS-ISPD2005 suboptimality benchmarks parametrized by white-space fraction. Displayed are quality ratios of total computed HPWL to near-optimal HPWL upper bounds. Results with uniformly distributed white space are shown for 5%, 10%, 20%, and the maximum possible white space values. For Peko-MS-adaptec1–4, quality ratios are also shown for benchmarks with optimal zero-white-space layouts ("pack") on the left side of the core region and 10% white space on the right. "mem" denotes an out-of-memory error. Capo 9.5 was run with option-noHMetis on Peko-MS-a3, Peko-MS-a4, and all four of the packed benchmarks; otherwise, all tools are in default mode in all cases.

| ckt\ ws | mPL6 | | | | | APlace 2.0 | | | | | Capo 9.5 | | | | |
|---|---|---|---|---|---|---|---|---|---|---|---|---|---|---|---|
| | pack | 5% | 10% | 20% | max | pack | 5% | 10% | 20% | max | pack | 5% | 10% | 20% | max |
| PWS-A1 | 1.80 | 1.35 | 1.48 | 1.70 | 1.80 | 1.33 | 1.50 | 1.22 | 1.15 | 1.54 | 6.17 | 3.33 | 3.14 | 3.05 | 2.67 |
| PWS-A2 | 2.11 | 1.48 | 1.48 | 1.36 | 1.54 | 3.46 | fail | fail | 3.65 | 2.29 | 8.06 | 4.01 | 3.85 | 3.53 | 3.12 |
| PWS-A3 | 4.32 | 2.14 | 1.52 | 1.41 | 1.33 | 2.27 | 1.23 | 1.14 | 1.13 | 1.10 | 4.10 | 2.10 | 1.93 | 1.53 | 1.33 |
| PWS-A4 | 4.39 | 1.50 | 1.32 | 1.51 | 1.24 | 1.70 | 1.29 | 1.23 | 1.34 | 1.44 | 3.09 | 2.08 | 1.92 | 1.62 | 1.35 |
| PWS-B1 | – | 1.30 | 1.34 | 1.30 | 1.24 | – | 1.44 | 1.32 | 1.17 | 1.33 | – | 2.50 | 2.42 | 2.08 | 1.77 |
| PWS-B2 | – | 2.10 | 2.16 | 1.64 | 1.39 | – | 1.25 | 1.26 | 1.58 | 1.44 | – | 2.42 | 2.13 | 1.83 | 1.51 |
| PWS-B3 | – | 1.54 | 1.62 | 1.99 | 2.02 | – | 2.14 | 1.59 | 2.02 | 2.23 | – | 2.49 | 2.10 | 1.89 | 1.91 |
| PWS-B4 | – | 1.51 | 1.46 | 1.71 | mem | – | 1.26 | 1.21 | 1.16 | 1.33 | – | mem | mem | mem | mem |
| Averages | 3.16 | 1.61 | 1.55 | 1.58 | 1.51 | 2.19 | 1.45 | 1.28 | 1.65 | 1.59 | 5.35 | 2.70 | 2.50 | 2.22 | 1.96 |

   Net-box corner-pin locations are selected at randomized distances from one another approximately $1/10$ of the width or height of the placement region, until the end terminal of the chain is reached.

Results of APlace 2.0, Capo 10, and mPL6, all run in default mode, are shown for the combined PEKO-MSPEKO-MC IBM01 benchmark in Table 2.3, both without and with nonlocal nets. Macros larger than ten cell rows high were treated as fixed, their boundaries thus supplying some additional terminal locations. As expected, the presence of monotone chains of nonlocal nets decreases all placers' suboptimality ratios.

**Table 2.3.** HPWL Suboptimality of APlace 2.0, Capo 10, and mPL6, compared on 10% and 40% white-space versions of a PWS circuit derived from the ICCAD 2004 IBM01 mixed-size benchmark, both without (*top*) and with (*bottom*) the addition of optimal-HPWL nonlocal nets. Approximately 13,15% of the nets in the second set are nonlocal, accounting for 57%, 68% of total HPWL.

| With Local Nets Only | | | |
|---|---|---|---|
| | APlace | Capo | mPL |
| ibm01-10WS | 1.20 | 1.88 | 1.31 |
| ibm01-40WS | 1.40 | 1.96 | 1.27 |
| Averages | 1.30 | 1.92 | 1.29 |

| With Chains of Optimal-HPWL Non-local Nets | | | | | |
|---|---|---|---|---|---|
| | $\frac{\#nln}{\#nets}$ | $\frac{WL_{nonloc}}{WL_{total}}$ | APlace | Capo | mPL |
| ibm01-10WS | 0.15 | 0.57 | 1.11 | 1.49 | 1.16 |
| ibm01-40WS | 0.13 | 0.68 | 1.08 | 1.67 | 1.10 |
| | | Averages | 1.10 | 1.58 | 1.13 |

### 2.4.4  Suboptimality of Detailed Placement

Optimal GPs (OGP) parametrized by bin size were generated from the optimal PEKO-MS placements as follows. Uniform rectangular bin grids of user-specified dimensions were superimposed. Cells and macros centered in the same bin were moved to the bin center, where they were placed concentrically. These OGP placements were then used as benchmarks for the DP engines of mPL6 [6, 11] and APlace2.0 [17]. Each PEKO-MS circuit can generate several different OGP circuits, one for each bin size. The DP engines were run on a set of these OGP circuits, and the rate of degradation in their quality with respect to bin size and white-space value was observed. For each of the different white-space values, the quality ratios obtained by the DP engines were averaged over the eight different circuits. The result is illustrated in Figure 2.9. The benchmarks reveal opposite trends in these engines with respect to increasing white space. For these test cases, mPL's performance degrades as white space increases, while APlace's improves. APlace's cell-swapping strategy may have some advantage on these benchmarks, because the standard cells in these test cases are all of uniform size and shape. Under higher white space, the size of the set of candidate swaps is reduced, making successful swaps more likely to be found. On the other hand, mPL's local-window-based refinement is apparently a drawback on the higher-white-space cases, where larger scale moves are apparently needed.

Results on the OGP benchmark derived from the PEKO-MS-adaptec2 benchmark with 10% uniformly distributed white space are summarized in Figure 2.10. Results are shown for two scenarios: one in which all macros are held fixed, and hence only standard cells are aggregated into bin centers, and another in which both kinds of objects are moved from their locations in the known near-optimal placement to the nearest bin center. The results of these experiments show that the quality of DP deteriorates fairly rapidly as the bin size increases, even for these uniformly accurate GPs. For bin sizes up to $4 \times 4$, macro legalization is not a primary source of suboptimality, but for larger bin sizes, it is.

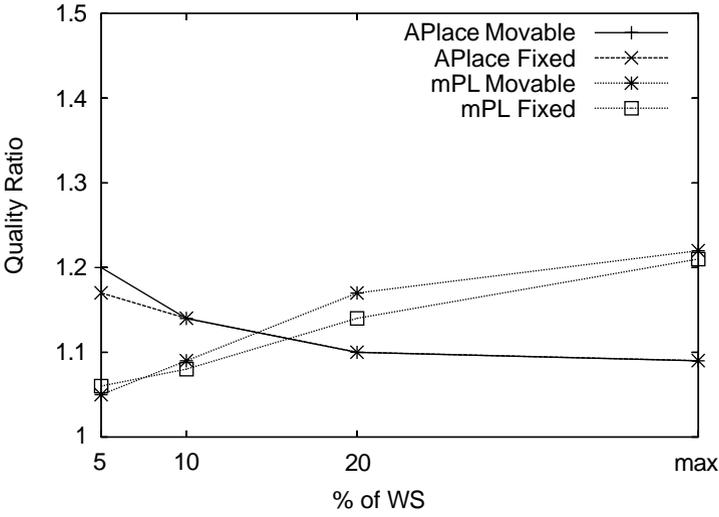

**Fig. 2.9.** Average quality ratios of APlace2.0-DP and mPL6-DP over the eight different netlists of OGP DP benchmarks.

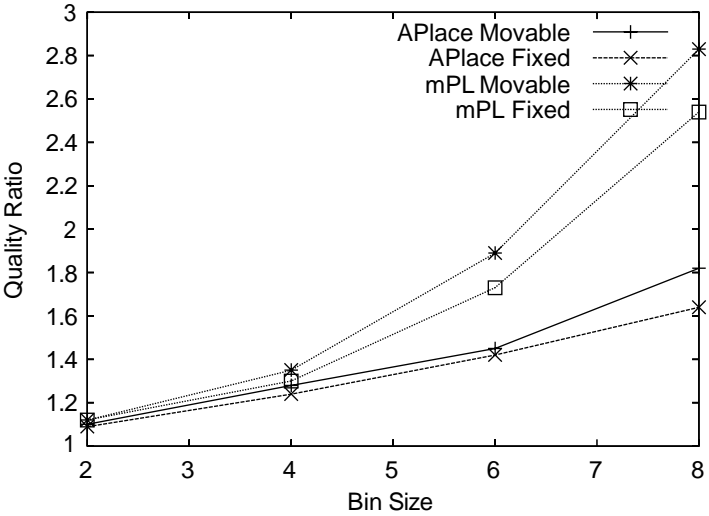

**Fig. 2.10.** Suboptimality of APlace2.0-DP and mPL6-DP on the OGP DP benchmarks generated from Peko-MS-adaptec2 with 10% white space.

**Table 2.4.** Estimated suboptimality of mPL6 global (GP) and detailed placement (DP) engines. The GP estimate is obtained simply by subtracting the observed DP quality ratio obtained on the $2 \times 2$ OGP benchmark from the overall quality ratio observed for the corresponding Peko-MS benchmark.

| %WS | GP | DP | Total |
|-----|-----|-----|-------|
| 5 | 51% | 10% | 61% |
| 10 | 46% | 9% | 55% |
| 20 | 39% | 17% | 56% |
| 40 | 29% | 22% | 51% |

The OGP benchmarks provide a means of estimating how much of a placer's suboptimality is attributable to its GP, and how much to legalization and DP. Table 2.4 compares the suboptimality observed for mPL on $2 \times 2$ OGP cases to that observed for mPL6, including both GP and DP, on the corresponding PEKO-MS source circuits from which the OGP cases are derived. Subtracting the observed DP suboptimality on the $2 \times 2$ OGP benchmark from the total mPL6 GP+DP suboptimality on the corresponding PEKO-MS benchmark gives an estimate of the mPL6 GP suboptimality. It should be noted, however, that these suboptimality values are not truly additive, for at least two reasons. First, the starting configuration for DP on the OGP benchmark is very different from the DP starting configuration on the corresponding PEKO-MS benchmark. Second, relative module positions in a GP below the resolution of the $2 \times 2$ OGP grid will typically be used as hints during legalization and DP to improve results.

Figure 2.11 displays line segments between modules' placed locations and their optimal locations in a small PEKO-MS testcase (IBM02) constructed with 5% white space from the ICCAD2004 mixed-size suite [1]. Results for both global and detailed placements of APlace 2.0 and mPL6 are shown. From these plots, it is clear that displacement errors are not at all randomly distributed and, on the contrary, display large-scale systematic bias. We observe similar trends in displacement plots for other tools on other Peko-MS circuits and at other white-space fractions. We conclude that, even when each cell in a GP is very close to (one of) its optimal location(s), further reduction in the objective can often only be achieved by moving large subsets of cells simultaneously by small amounts. Iterative, local, window-based refinement will not remove the systematic error.

### 2.4.5 HPWL Suboptimality Comparison of Leading Academic Tools on Peko-MS 2005

Recent (early 2007) versions of the placers entered in the 2006 ISPD Placement Contest were run on PEKO-MS adaptations of the ISPD 2005 and ISPD 2006 benchmark suites. The ratios of attained HPWL to the known near-optimal HPWL on the 2005 PEKO-MS benchmarks at 80% free-space utilization are shown in Table 2.5. Summary statistics for these benchmarks are shown in Table 2.6. Notational abbreviations are summarized in Table 2.7.

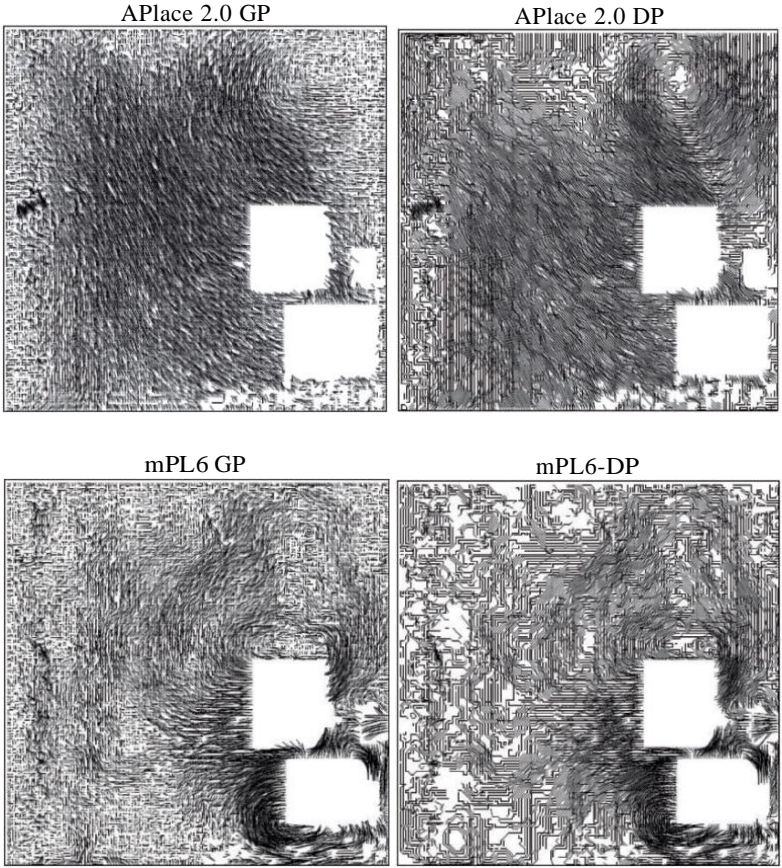

**Fig. 2.11.** Individual module displacements from optimal on the Peko-MS-ICCAD04-IBM02 benchmark with 5% white space. Displacements of both global and detailed placements are shown for APlace (*top*) and mPL6-DP (*bottom*). The HPWL quality ratios observed on this benchmark are 1.45 for APlace and 1.23 for mPL.

There are at least two ways in which these results may be useful in identifying weaknesses of the tools. First, the PEKO-MS benchmarks tend to amplify the suboptimality associated with local nets. Hence, a relatively high-average suboptimality gap on these test cases by a given tool (e.g., DPlace, mFar) suggests that tool might benefit from enhancements designed to help it better identify such local nets and reduce their lengths. Second, detailed investigation of a given tool's computation on a particular test case where it exhibits a relatively large gap (e.g., Kraftwerk on bigblue2, APlace on bigblue1, mPL6 on bigblue3) compared to its own results on other test cases may be useful in improving the tool's robustness. Such anomalous gaps may be particularly useful when the circuits on which they are observed have some distinguishing features, e.g., the relatively large number of movable macros in bigblue2, or the relatively low-area-fraction of fixed objects in bigblue1.

**Table 2.5.** HPWL suboptimality ratios of leading academic placers on Peko-MS ISPD 2005 benchmarks. For notation, see Table 2.7.

| | PEKO-MS-05 (80% free-space utilization) suboptimality ratios | | | | | | | | | |
|---|---|---|---|---|---|---|---|---|---|---|
| Circuit | DPlace | Kraft | Capo | APlace | FastP | Mfar | NTUPlace3 | mPL6 | Dragon | Averages |
| adaptec1 | 1.47 | 1.19 | 1.50 | 1.13 | 1.58 | 2.44 | 1.30 | 1.27 | 2.63 | 1.61 |
| adaptec2 | 1.62 | 1.17 | 1.61 | 1.12 | 1.61 | 2.61 | 1.61 | 1.32 | 3.07 | 1.75 |
| adaptec3 | 1.77 | 1.21 | 1.60 | 1.13 | 1.79 | 2.50 | 1.58 | 1.43 | 2.95 | 1.77 |
| adaptec4 | 1.73 | 1.23 | 1.47 | 1.13 | 1.71 | 2.21 | 1.37 | 1.29 | 2.21 | 1.59 |
| bigblue1 | 1.52 | 1.22 | 1.42 | 1.31 | 1.63 | 2.67 | 1.24 | 1.21 | 2.74 | 1.66 |
| bigblue2 | 1.61 | 1.45 | 1.57 | 1.29 | 1.65 | – | 1.34 | 1.34 | 4.75 | 1.87 |
| bigblue3 | 2.49 | 1.27 | 2.01 | 1.19 | 1.84 | 3.04 | 3.01 | 1.56 | 6.73 | 2.57 |
| bigblue4 | 2.20 | 1.29 | 1.48 | 1.29 | 2.00 | 3.61 | 1.39 | 1.31 | 3.03 | 1.96 |
| Averages | 1.80 | 1.25 | 1.58 | 1.20 | 1.73 | 2.72 | 1.60 | 1.34 | 3.51 | 1.85 |

**Table 2.6.** Peko-MS ISPD 2005 benchmarks statistics. For notation, see Table 2.7.

| | PEKO-MS-ISPD2005 statistics (80% free-space utilization) | | | | | | | | | |
|---|---|---|---|---|---|---|---|---|---|---|
| Circuit | #obj | #mac | a(mac) | a(fix) | #term | # net | # pin | #pad | util | HPWL |
| adaptec1 | 211977 | 63 | 0.431 | 0.430 | 542 | 243402 | 939918 | 480 | 0.80 | 20056216 |
| adaptec2 | 261153 | 159 | 0.615 | 0.613 | 543 | 290584 | 1061117 | 407 | 0.80 | 24969764 |
| adaptec3 | 466560 | 723 | 0.615 | 0.615 | 723 | 515720 | 1864504 | 0 | 0.80 | 40954784 |
| adaptec4 | 511448 | 1129 | 0.486 | 0.486 | 1329 | 558781 | 1903080 | 0 | 0.80 | 39391712 |
| bigblue1 | 273708 | 32 | 0.172 | 0.172 | 559 | 307892 | 1136899 | 528 | 0.80 | 20858240 |
| bigblue2 | 578938 | 22984 | 0.384 | 0.346 | 3313 | 634210 | 2117165 | 0 | 0.80 | 42256768 |
| bigblue3 | 1122682 | 3778 | 0.680 | 0.668 | 675 | 1207133 | 3791237 | 0 | 0.80 | 94399040 |
| bigblue4 | 2244173 | 8169 | 0.376 | 0.358 | 664 | 2494307 | 8792369 | 0 | 0.80 | 171477120 |

## 2.4.6 Suboptimality of Routability-Aware Placement

The Peko-MS algorithm was adapted to model the 2006 ISPD placement-contest scaled-HPWL objective, excluding run time, as follows. The scaled HPWL objective is computed as $HPWL \times (1 + 0.01 \, \sigma)$, where $\sigma$ is the "scaled overflow factor," defined by summation of bin-overflow penalties over each bin in a uniform grid consisting of square bins ten standard-cell rows high. We refer to this uniform grid as the "utilization evaluation grid." The utilization penalty for a given bin B is [20],

$$\sum_{\{mod\,ules\,m\}} \text{area of } m \text{ overlapping with B} - \text{utilization-target} \times \text{free area(B)},$$

where the free area in B is simply the total area in B not occupied by fixed objects. Although this measure may vary somewhat with the size and position of the utilization-evaluation grid relative to fixed objects, it does test a placer's ability to target area congestion in a specified subregion.

The PEKO-MS generator of Figure 2.2 in Sect. 2.3 was adapted to target user-specified bin utilizations simply by iterating its random white-space insertion separately over all bins in the utilization-evaluation grid, terminating when either the utilization target is reached or when a given limit on BFS iterations (from 200 to 800) is reached. Complex fixed-macro geometries often make the precise target difficult for this simple approach to attain; hence, the final bin-utilization Peko-MS benchmarks have a low but nonzero and hence suboptimal level of overflow in most evaluation bins. Characteristics of these benchmarks are described in Table 2.8.

**Table 2.7.** Notation for column labels in this section.

| Notation | |
|---|---|
| #obj | number of objects, movable and fixed |
| #mac | number of macros, i.e., objects of height greater than 1 row |
| a(mac) | fraction of core area occupied by macros, movable, or fixed |
| a(fix) | fraction of core area occupied by fixed objects |
| #term | number of fixed terminals |
| # net | number of nets |
| # pin | number of pins |
| #pad | number of perimeter I/O objects |
| util | target utilization in the part of core not occupied by fixed objects |
| HPWL | total HPWL of the given near-optimal placement |
| Hratio | ratio of attained HPWL to near-optimal HPWL |
| SOV/bin | average scaled overflow per bin |
| SHPWL | HPWL, scaled by $1 + 0.01 \times$SOV/bin |
| Sratio | ratio of attained SHPWL to near-optimal SHPWL |

**Table 2.8.** Peko-MS ISPD 2006 benchmarks statistics. For notation, see Table 2.7.

| PEKO-MS-06 statistics | | | | | | | |
|---|---|---|---|---|---|---|---|
| Circuit | # objs | # macs | a(mac) | a(fix) | #term | # nets | # pins |
| adaptec5 | 872276 | 646 | 0.572 | 0.572 | 646 | 1029763 | 3394072 |
| newblue1 | 385625 | 64 | 0.379 | 0.000 | 337 | 334324 | 1237412 |
| newblue2 | 461252 | 5000 | 0.652 | 0.636 | 1171 | 463382 | 1772264 |
| newblue3 | 511413 | 8756 | 0.792 | 0.792 | 8845 | 559874 | 1940730 |
| newblue4 | 671548 | 3422 | 0.359 | 0.359 | 3422 | 711993 | 2418450 |
| newblue5 | 1282550 | 4881 | 0.495 | 0.495 | 4881 | 1520814 | 4805020 |
| newblue6 | 1318990 | 6505 | 0.334 | 0.334 | 6505 | 1301252 | 5305156 |
| newblue7 | 2641754 | 25065 | 0.535 | 0.535 | 25065 | 2651867 | 10098844 |

| Circuit | #pads | util (%) | hpwl | sov/bin | shpwl |
|---|---|---|---|---|---|
| adaptec5 | 0 | 50 | 81893792 | 9.99 | 9.01E+07 |
| newblue1 | 337 | 80 | 20500032 | 1.73 | 2.09E+07 |
| newblue2 | 0 | 90 | 32869280 | 10.29 | 3.63E+07 |
| newblue3 | 0 | 80 | 73514272 | 9.55 | 8.05E+07 |
| newblue4 | 0 | 50 | 49143584 | 9.26 | 5.37E+07 |
| newblue5 | 0 | 50 | 102083104 | 9.58 | 1.12E+08 |
| newblue6 | 0 | 80 | 90657856 | 8.36 | 9.82E+07 |
| newblue7 | 0 | 80 | 206175072 | 7.07 | 2.21E+08 |

**Table 2.9.** HPWL and SHPWL suboptimality ratios of leading academic placers on Peko-MS ISPD-2006 benchmarks. For notation, see Table 2.7.

| PEKO-MS-06 suboptimality ratios | | | | | | | | |
|---|---|---|---|---|---|---|---|---|
| | DPlace | | | Kraftwerk | | | Capo | | |
| Circuit | Hratio | SOV/bin | Sratio | Hratio | SOV/bin | Sratio | Hratio | SOV/bin | Sratio |
| adaptec5 | 2.53 | 369.8 | 10.80 | 1.17 | 36.37 | 1.45 | 1.58 | 4.97 | 1.51 |
| newblue1 | 2.42 | 116.1 | 5.14 | 1.49 | 11.6 | 1.63 | 2.75 | 1.53 | 2.74 |
| newblue2 | 2.18 | 124.2 | 4.42 | 1.29 | 50.87 | 1.76 | 2.77 | 1.17 | 2.54 |
| newblue3 | 2.19 | 141.8 | 4.82 | 1.10 | 53.27 | 1.55 | 1.65 | 1.72 | 1.53 |
| newblue4 | 2.22 | 282.5 | 7.77 | 1.23 | 36.26 | 1.54 | 1.61 | 6.43 | 1.57 |
| newblue5 | 2.96 | 360.7 | 12.45 | 1.37 | 66.94 | 2.09 | 1.64 | 6.33 | 1.60 |
| newblue6 | 2.28 | 168.5 | 5.66 | 1.23 | 51.88 | 1.72 | 2.15 | 2.3 | 2.03 |
| newblue7 | 3.70 | 261.0 | 12.48 | 1.25 | 46.07 | 1.71 | 2.04 | 2.11 | 1.94 |
| Averages | 2.56 | 228.1 | 7.94 | 1.27 | 44.16 | 1.68 | 2.02 | 3.32 | 1.93 |
| | APlace | | | FastPlace | | | MFar | | |
| Circuit | Hratio | SOV/bin | Sratio | Hratio | SOV/bin | Sratio | Hratio | SOV/bin | Sratio |
| adaptec5 | 1.13 | 117.2 | 2.23 | 2.09 | 100.3 | 3.80 | 3.30 | 213.1 | 9.40 |
| newblue1 | 1.35 | 89.7 | 2.51 | 2.18 | 13.3 | 2.43 | 5.98 | 39.1 | 8.18 |
| newblue2 | 1.43 | 162.5 | 3.41 | 1.61 | 45.4 | 2.12 | 3.72 | 186.8 | 9.66 |
| newblue3 | 1.20 | 132.6 | 2.55 | 1.11 | 81.3 | 1.84 | 2.15 | 162.8 | 5.15 |
| newblue4 | 1.12 | 75.1 | 1.80 | 1.54 | 98.5 | 2.80 | 3.08 | 219.2 | 9.00 |
| newblue5 | 2.61 | 217.6 | 7.55 | 2.05 | 84.8 | 3.46 | 3.16 | 207.7 | 8.87 |
| newblue6 | 1.15 | 48.4 | 1.57 | 1.39 | 45.6 | 1.87 | 2.98 | 176.5 | 7.59 |
| newblue7 | 1.31 | 119.8 | 2.68 | 1.32 | 45.1 | 1.79 | 2.74 | 177.1 | 7.10 |
| Averages | 1.41 | 120.34 | 3.04 | 1.66 | 64.28 | 2.51 | 3.39 | 172.78 | 8.12 |
| | NTUPlace3 | | | mPL6 | | | Dragon | | |
| Circuit | Hratio | SOV/bin | Sratio | Hratio | SOV/bin | Sratio | Hratio | SOV/bin | Sratio |
| adaptec5 | 1.31 | 7.4 | 1.28 | 1.35 | 17.7 | 1.44 | 2.96 | 0.29 | 2.69 |
| newblue1 | 1.26 | 14.2 | 1.42 | 1.50 | 11.5 | 1.65 | 3.11 | 0.01 | 3.06 |
| newblue2 | 1.45 | 2.6 | 1.35 | 1.35 | 31.5 | 1.61 | 4.20 | 0.13 | 3.81 |
| newblue3 | 1.28 | 5.3 | 1.23 | 1.36 | 20.6 | 1.50 | 3.49 | 0.14 | 3.19 |
| newblue4 | 1.26 | 3.3 | 1.19 | 1.37 | 15.8 | 1.45 | 2.64 | 0.29 | 2.43 |
| newblue5 | 1.29 | 4.7 | 1.23 | 1.29 | 17.0 | 1.38 | 2.94 | 0.30 | 2.69 |
| newblue6 | 1.23 | 2.0 | 1.16 | 1.40 | 18.0 | 1.53 | 2.80 | 0.17 | 2.59 |
| newblue7 | 1.36 | 8.3 | 1.38 | 1.54 | 26.7 | 1.83 | 3.76 | 0.11 | 3.52 |
| Averages | 1.30 | 6.0 | 1.28 | 1.40 | 19.8 | 1.55 | 3.24 | 0.18 | 3.00*** |

Results of the most recent available implementations of the placers entered in the ISPD 2006 Placement Contest on the 2006 PEKO-MS test cases are shown in Table 2.9; median results over all the placers are listed in Table 2.10. Overall, the high-scaled HPWL suboptimality values obtained by most tools on most of the these benchmarks reveals considerable room for improvement of these tools in the presence of congestion metrics. As with the other PEKO-MS benchmarks, the utility of the 2006 PEKO-MS test cases lies primarily in helping to identify particular test cases where investigation of a given tool's performance may reveal weaknesses.

**Table 2.10.** Median HPWL and SHPWL suboptimality ratios of all leading academic placers listed in Table 2.9 on Peko-MS ISPD-2006 benchmarks. For notation see Table 2.7.

| Median PEKO-MS-06 Suboptimality Ratios | | | |
|---|---|---|---|
| Circuit | Hratio | SOV/bin | Sratio |
| adaptec5 | 1.58 | 36.37 | 2.23 |
| newblue1 | 2.18 | 13.27 | 2.51 |
| newblue2 | 1.61 | 45.40 | 2.54 |
| newblue3 | 1.36 | 53.27 | 1.84 |
| newblue4 | 1.54 | 36.26 | 1.80 |
| newblue5 | 2.05 | 66.94 | 2.69 |
| newblue6 | 1.40 | 45.61 | 1.87 |
| newblue7 | 1.54 | 45.06 | 1.94 |
| Medians | 1.56 | 45.23 | 2.08 |

E.g., consider (1) Capo on newblue1 and newblue2, and (2) APlace, DPlace, and FastPlace on newblue5, etc. While such hints might also be obtained simply by comparing to results of several tools on the original ISPD 2006 benchmarks, use of the PEKO-MS test cases may reduce the time needed to identify deficiencies by providing an absolute measure of suboptimality. In particular, the PEKO-MS test cases facilitate analysis of the trade-off between routability optimization and HPWL optimization without the need for comparisons to results of other tools. E.g., on the PEKO-MS test cases, results suggest that the superior area-congestion reduction of Capo and Dragon comes at a significant cost in increased HPWL. APlace, on the other hand, typically attains excellent HPWL reduction but relatively high-bin-overflow values; this result suggest that its placements may sometimes be difficult to route.

## 2.5 Conclusions

Two new sets of synthetic benchmark circuits with known optimal-HPWL or near-optimal-HPWL placements have been presented. The PEKO-MC set quantifies the role of nonlocal nets in suboptimality; the PEKO-MS set quantifies the role of white space and modules of mixed size. Experiments with leading academic placement tools support four main conclusions. First, as shown in Table 2.2, different tools produce widely varying results on some of the mixed-size PEKO-MS benchmarks. Hence, these benchmarks can be used to identify deficiencies in tools producing relative poor results. Second, the presence of netwise-disjoint chains of nets linking pairs of numerous, well distributed, fixed terminals appears to make wire length-driven placement by contemporary methods considerably less difficult. Circuits designed to ensure the existence of monotone paths for all signals [23, 24, 26, 27] might reasonably be expected to have wire lengths far closer to optimal than what leading placement tools are able to achieve on other circuits. Third, the accumulation of small but systematic errors in the placement of local nets appears to be a greater source of suboptimality than the total error in identifying and placing nonlocal nets.

The corrective action needed to further reduce that suboptimality, whether taken during global placement, legalization, or detailed placement, must consider simultaneous motion of large subsets of objects in order to be effective. Restriction to subsets localized in an arbitrary way is, in general, insufficient to improve on existing results. Fourth, the high-scaled HPWL suboptimality values obtained by most tools on most of the bin-utilization-controlled PEKO-MS benchmarks suggest that considerable room for improvement of these tools remains, particularly on large complex test cases, and particularly when bin-area congestion is factored into the quality evaluation.

## 2.6 Acknowledgments

Financial support for this work was provided by Semiconductor Research Consortium Contract 2003-TJ-1091 and National Science Foundation Contracts CCF 0430077 and CCF-0528583.